\documentclass[prl,twocolumn,superscriptaddress,showpacs,floatfix]{revtex4}

\usepackage{graphicx}
\usepackage{SIunits}
\usepackage[normal]{subfigure}
\usepackage{color}
\usepackage{amsmath} 

\newcommand{\comments}[1]{}

\begin{document}

\title{Efficient out-coupling of high-purity single photons from a coherent quantum dot in a photonic-crystal cavity}

\author{K.~H.~Madsen}\email{khmadsen@nbi.ku.dk}
\affiliation{Niels Bohr Institute, University of Copenhagen, Blegdamsvej 17, DK-2100 Copenhagen, Denmark}
\affiliation{DTU Fotonik, Department of Photonics Engineering, Technical University of Denmark, \O rsteds Plads 343, DK-2800 Kgs.\ Lyngby, Denmark}
\author{S.~Ates}
\altaffiliation[Present address: ]{The National Research Institute of Electronics and Cryptology,
The Scientific and Technological Research Council of Turkey (TUBITAK), Gebze 41400, Turkey}
\affiliation{DTU Fotonik, Department of Photonics Engineering, Technical University of Denmark, \O rsteds Plads 343, DK-2800 Kgs.\ Lyngby, Denmark}
\author{J.~Liu}
\affiliation{Niels Bohr Institute, University of Copenhagen, Blegdamsvej 17, DK-2100 Copenhagen, Denmark}
\affiliation{DTU Fotonik, Department of Photonics Engineering, Technical University of Denmark, \O rsteds Plads 343, DK-2800 Kgs.\ Lyngby, Denmark}
\author{A.~Javadi}
\affiliation{Niels Bohr Institute, University of Copenhagen, Blegdamsvej 17, DK-2100 Copenhagen, Denmark}
\author{S.~M.~Albrecht}
\affiliation{Niels Bohr Institute, University of Copenhagen, Blegdamsvej 17, DK-2100 Copenhagen, Denmark}
\author{I.~Yeo}
\affiliation{Niels Bohr Institute, University of Copenhagen, Blegdamsvej 17, DK-2100 Copenhagen, Denmark}
\author{S.~Stobbe}
\affiliation{Niels Bohr Institute, University of Copenhagen, Blegdamsvej 17, DK-2100 Copenhagen, Denmark}
\author{P.~Lodahl}\email{lodahl@nbi.ku.dk} \homepage{www.quantum-photonics.dk}
\affiliation{Niels Bohr Institute, University of Copenhagen, Blegdamsvej 17, DK-2100 Copenhagen, Denmark}

\date{\today}

\begin{abstract}
We demonstrate a single-photon collection efficiency of $(44.3\pm2.1)\%$ from a quantum dot in a low-Q mode of a photonic-crystal cavity with a single-photon purity of $g^{(2)}(0)=(4\pm5)\%$ recorded above the saturation power. The high efficiently is directly confirmed by detecting up to $962\pm46$ kilocounts per second on a single-photon detector on another quantum dot coupled to the cavity mode. The high collection efficiency is found to be broadband, as is explained by detailed numerical simulations. Cavity-enhanced efficient excitation of quantum dots is obtained through phonon-mediated excitation and under these conditions, single-photon indistinguishability measurements reveal long coherence times reaching $0.77\pm0.19$ ns in a weak excitation regime. Our work demonstrates that photonic crystals provide a very promising platform for highly integrated generation of coherent single photons including the efficient outcoupling of the photons from the photonic chip.

\end{abstract}

\maketitle

The ability to reduce decoherence processes of quantum dots (QDs) is of high importance for their utilization in quantum-information processing~\cite{Knill.Nature.2001}, where indistinguishable and on-demand single photons are highly desirable~\cite{Lodahl.arXiv.2013}. The influence of dephasing on the quantum interference between consecutively emitted photons from a single QD has been studied previously~\cite{Santori.Nature.2002,Laurent.APL.2005}, where it was suppressed by Purcell enhancing the emitter decay rate. Dephasing is partly attributed to spectral diffusion arising from fluctuations in both the electrostatic environment and in the nuclear spin ensemble~\cite{Kuhlmann.arXiv.2013}. Such processes give rise to fluctuating electric and magnetic fields on the timescale of $\gtrsim10\;\milli$s and $\gtrsim10\;\micro$s, respectively, but these are much slower than the nanosecond timescale of the QD dynamics that is relevant for the generation of indistinguishable photons. On the other hand, the interaction between the exciton and longitudinal acoustic (LA) phonons is an important fast dephasing mechanism with a characteristic timescale of picoseconds, which gives rise to broad sidebands that can be spectrally filtered with cavities~\cite{Kaer.PRB.2013}. Recently, nearly perfectly indistinguishable photons were demonstrated by the use of pulsed resonant excitation~\cite{He.NatNan.2013}, although so far this was not implemented in photonic nanostructures, which limits the efficiency. The current state of the art for optimizing both indistinguishability and efficiency using non-resonant excitation schemes was reported in Ref.~\cite{Gazzano.NatCom.2013} for the case of a micropillar cavity.

In the present work we report on measurements on QDs spectrally close to a low-Q mode of a photonic-crystal (PC) cavity. We show that the QD emission can be very efficiently collected by a microscope objective and derive a collection efficiency of $(44.3 \pm2.1)\%$ at the first lens by comparing to a QD situated in an unprocessed region of the wafer (referred to as bulk GaAs).
The collection efficiency is defined as the average number of photons hitting the first lens divided by the average number of photons emitted from the QD per excitation pulse.
Even when the QD is driven above saturation, the emission from the QD remains anti-bunched. The expected collection efficiency is calculated by simulating the far-field emission profile of the QD in the cavity. The simulations reveal that the high collection efficiency is expected to be very broadband, which is verified experimentally. In an effort to optimize directly the number of detected single photons, we record in a high-throughput optical setup a count rate of $(962\pm46)$ kHz on an avalanche photodiode (APD) from a QD exhibiting anti-bunching. In most applications of the single-photon source, e.g., for linear optics quantum computing~\cite{Kok.Rev.2007}, the actual rate of detected photons is decisive rather than the inferred collection efficiency that is often reported in the literature.  The high count rate achievable with QD sources underlines their potential for quantum-information applications, and indeed the first proof-of-concept demonstrations have recently emerged~\cite{Gazzano.PRL.2013}.

The excitation conditions are known to have a decisive impact on the coherence of the photons emitted from QDs. We demonstrate phonon-mediated excitation, where the QD is excited through a longitudinal-optical (LO) phonon that is resonant with a high-order mode of the cavity  or through LA phonon side-band. The indistinguishability of consecutively emitted photons from the QD is measured under both LO- and LA-phonon-mediated excitation. Although the QD decay rate is inhibited due to the photonic band-gap effect, a pronounced degree of indistinguishability is observed, which implies that the dephasing rate is low.
These low dephasing rates prove that the PC platform is well suited for the generation of highly coherent single photons.

\section{Single-photon collection efficiency}

\begin{figure*}[tb]
\includegraphics[width=13.3cm]{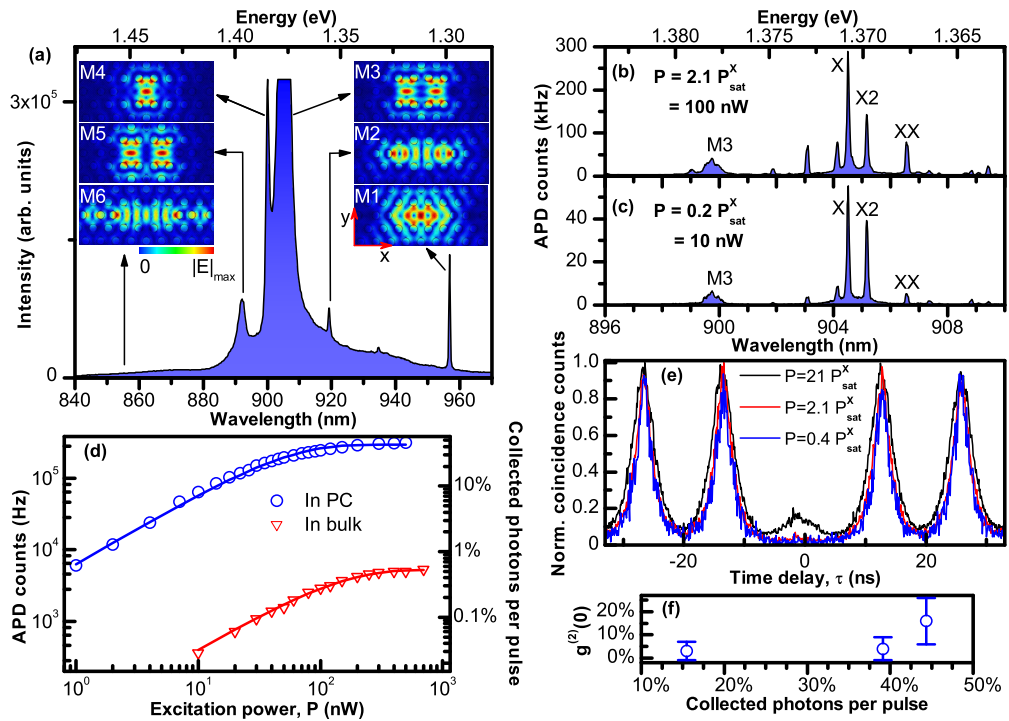}
\caption{Measurements on QD1 in Setup 1. \textbf{(a)} Cavity emission spectrum showing the modes labeled M1-M6 under strong above-band CW excitation. The emission from the M3 mode saturates the CCD-camera. Insets show the magnitude of the simulated electric fields of the M1-M6 modes. \textbf{(b)-(c)} Emission spectra recorded with an APD in a spectral range close to M3 for high and low power under pulsed M6-resonant excitation. The neutral exciton of interest (X), charged exciton (X2), and bi-exciton (XX) are identified using cross-correlation measurements and power series. \textbf{(d)} Power series of the X-line and of the emission from a QD in bulk GaAs recorded on the APD under pulsed excitation together with their fits (solid lines). The count rate of the X-line above saturation is $2.93\cdot 10 ^5\; \mathrm{counts \; s^{-1}}$, which is a $56$-fold increase compared to the QD in bulk GaAs corresponding to a collection efficiency at the first lens of $(44.3\pm 2.1)\;\%$. \textbf{(e)} Auto-correlation measurements of the X-line for three different excitation powers showing very pronounced anti bunching, and in \textbf{(f)} the extracted $g^{(2)}(\tau=0)$-values are plotted as a function of collected photons per excitation pulse.
\label{fig:1}}
\end{figure*}

We investigate an L3 PC cavity~\cite{Akahane.OE.2005} containing InGaAs QDs and select a single polarization of the emission. Details on the sample and setup can be found in Appendix A. In Fig.~\ref{fig:1}(a) the emission spectrum of the cavity is shown under strong above-band ($800$ nm) continuous wave (CW) excitation. The modes of the cavity are labeled M1-M6 with M1 being the first-order high-Q mode. The QD ensemble emits within the region $880-1000$ nm meaning that the M6 mode is not visible, but its presence can be revealed by scanning the excitation laser across it. Insets in Fig.~\ref{fig:1}(a) show the magnitude of the simulated electric fields of the M1-M6 cavity modes. M1, M2, M4, and M6 are all polarized along the y-axis in the far-field, while M3 and M5 have the orthogonal polarization in the far-field. Notably the M3 mode is observed to be very intense despite being a low-Q mode, which has also been reported previously~\cite{Oulton.OE.2007}, and this illustrates that light can be efficiently out-coupled from this mode.

In Fig.~\ref{fig:1}(b)-(c) high- and low-power spectra recorded on an APD are shown for the spectral range close to the M3 mode under pulsed excitation, while the excitation laser is tuned into resonance with the M6 mode~\cite{Kaniber.NJP.2009}. Using power series and cross-correlation measurements the neutral exciton (X), charged exciton (X2), and bi-exciton (XX) are identified for the emitter referred to as QD1. In Fig.~\ref{fig:1}(d) a power series of the X-line is shown, where the very high count rate on the APD signals a high collection efficiency. The power series is modeled with the function
C$_{\mathrm{out}}=$C$_{\mathrm{sat}} \left(1-\exp{\left(-\mathrm{P}/\mathrm{P}_{\mathrm{sat}}\right)} \right)$, cf. Fig.~\ref{fig:1}(d), where P and P$_{\mathrm{sat}}$ are the input power and saturation power measured before the microscope objective, and C$_{\mathrm{out}}$ and C$_{\mathrm{sat}}$ are the recorded count rates on the APD. We obtain P$^\mathrm{X}_{\mathrm{sat}}=46.7 \pm 3.7 \;\mathrm{nW}$ and C$^\mathrm{X}_{\mathrm{sat}}=(2.93\pm0.086)\cdot 10 ^5 \; \mathrm{counts \; s^{-1}}$ for the single X-line. Fig.~\ref{fig:1}(d) also contains a power series on a QD situated in bulk GaAs, for which we extract P$^{\mathrm{bulk}}_{\mathrm{sat}}=126.7\pm6.3 \;\mathrm{nW}$ and C$^{\mathrm{bulk}}_{\mathrm{sat}}=(5.22\pm0.10)\cdot 10 ^3\; \mathrm{counts \; s^{-1}}$. By comparing the two count rates, we conclude that the X-line from the QD in the PC cavity is $56$ times more efficiently collected than the excitonic emission from the QD in bulk GaAs.

Finite-element calculations of the collection efficiency of a QD in bulk GaAs and for a numerical aperture (NA) of $0.6$ for the collection lens gives an expected collection efficiency of $\eta_{\mathrm{bulk}}=0.79 \;\%$, which is in agreement with previous results in the literature~\cite{Claudon.NatPhot.2010}. For the QD in bulk GaAs we express the count rate on the APD at saturation as
\begin{align}\label{count_rate_exp}
C_\mathrm{sat}&=\eta_\mathrm{setup} \eta \mathrm{r} \;\;, \; \\
\mathrm{r}&=\alpha \epsilon \frac{\Gamma_\mathrm{l}}{2} \;\;.
\end{align}
$\eta_\mathrm{setup}$ is the probability of detecting a photon once a photon has been collected by the first lens, $\eta$ is the collection efficiency into the first lens, and $\mathrm{r}$ is the repetition rate of single-photon emission from the QD. The latter is linked to the repetition rate of the excitation laser, $\Gamma_\mathrm{l}$, where in the ideal case $\mathrm{r}=\Gamma_\mathrm{l}/2$ when driving the QD into saturation, where the factor of two expresses that a non-resonant excitation pulse can excite either of two orthogonally polarized dipoles in the QD while only one polarization component is detected. However, for two reasons this does not hold in general. Firstly, $x$- and $y$-oriented dipoles can both couple to the same linear polarization in the far field, and $\alpha$ denotes this degree of polarization mixing. This mixing is determined by the  position of the QD since the local polarization in the photonic-crystal structure generally varies strongly. We have  $0\leq \alpha \leq2$, where $\alpha=0$ ($\alpha=2$) corresponds to both dipoles emitting into the blocked (selected) far-field polarization component determined by the polarizer in the experiment.  $\alpha=1$ corresponds to no polarization mixing. Secondly, not every excitation of the QD actually gives rise to photon emission from one of the two neutral excitons, and $\epsilon$ denotes this photon-generation efficiency. We expect $\epsilon < 1$ since neutral exciton transitions in QDs may suffer from various charge-trapping processes that all would be relevant for the total photon generation efficiency. For instance, blinking processes between the neutral exciton in the QD and either charged excitons~\cite{Santori.PRB.2004} or dark excitons~\cite{Johansen.PRB.2010} may occur on a submicrosecond timescale that can effectively decrease the quantum efficiency of the QD~\cite{Johansen.PRB.2008}. Defect sites in the vicinity of the QD is another potential source of blinking that can be slower than the decay of the QD. Some of the blinking processes can be monitored in pulsed autocorrelation measurements by recording the variations of the peak amplitude at large time delays~\cite{Santori.PRB.2004,Davanco.arXiv.2013}. In the present experiment no evidence of blinking was found in correlation measurements up to $10$ ms for the QD in the PC cavity, but similar measurements for the QD in bulk GaAs were not possible due to the low count rate in this case. For a QD in bulk GaAs, simulations confirm that $\alpha_\mathrm{bulk}=1$ and under the idealized assumption of $\epsilon_\mathrm{bulk}=1$, an overall setup efficiency of $\eta_\mathrm{setup}=1.7\; \%$ is obtained. The collection efficiency for the  QD in the PC cavity can be obtained from Eq.~(\ref{count_rate_exp}) under the assumption that polarization mixing and photon-generation efficiencies are identical for the QD in the PC cavity and in bulk GaAs, i.e., $\mathrm{r_{X}}=\mathrm{r_{bulk}}.$ This leads to an extracted collection  efficiency of $\eta_\mathrm{X}=(44.3\pm 2.1) \; \%$ for the X-line. The potential influence of $\alpha$ and $\epsilon$ on this number is discussed in further detail below.

Time-resolved measurements enable determining the coupling efficiency to the cavity. The decay of the X-line is bi-exponential as expected for a neutral exciton~\cite{Johansen.PRB.2010} with a fast decay rate of $0.62$ ns$^{-1}$ and a slow decay rate of $0.24$ ns$^{-1}$. Compared to a QD in bulk GaAs the decay rate of the X-line is slightly inhibited. Nonetheless the QD is in fact significantly Purcell enhanced by the M3 cavity mode, since the 2D photonic band gap suppresses the coupling to radiation modes strongly~\cite{Wang.PRL.2011}. In Fig.~\ref{fig:1}(e) autocorrelation measurements of the X-line are shown for three very different excitation powers. We obtain g$^{(2)}(\tau=0)$ by integrating all counts in a 2 ns window around zero delay and dividing by the corresponding average area of the coincidence peaks observed in a $300$ ns window. We obtain the values g$^{(2)}(0)=(3 \pm 4)\%$, $(4 \pm 5)\%$, and $(16 \pm 10)\%$ for excitation powers of $0.4$, $2.1$, and $21$ times the saturation power, respectively. These values are shown in Fig.~\ref{fig:1}(f) as a function of collected photons per excitation pulse. Even at $21$ times saturation power, the anti bunching is very pronounced. The single-photon nature of the emission combined with the very high collection efficiency of up to $\eta_\mathrm{X}=(44.3\pm2.1)\%$ shows that PCs can be used also for vertically coupling photons out for immediate applications despite the fact that the platform is planar and therefore particularly suited for integration.

\begin{figure}[tb]
\includegraphics[width=8.3cm]{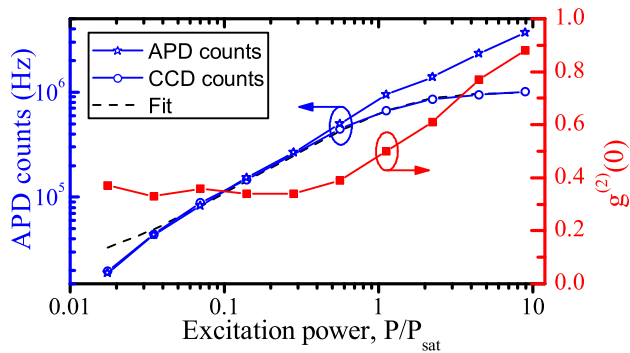}
\caption{Power series on QD2 using Setup 2 comparing measurements done for weak filtering with a band-pass filter and detection with an APD or strong filtering with a grating spectrometer and a CCD camera. The power-dependent $g^{(2)}(0)$ measured with the band-pass filter is also shown. The recorded counts on the CCD have been scaled to match the direct measurements on the APD in the limit of weak excitation. The saturation behavior of the CCD measurements is modeled as an exponential increase (dashed line, see main text for details) and from that a saturation count rate of  $\mathrm{C^X_{sat}}=962\pm 46\; \mathrm{kHz}$ is obtained.
\label{fig:Sven}}
\end{figure}
As an alternative and more direct measurement of the single-photon efficiency we next record the transmission throughput of the entire optical path. In an effort to achieve a higher count rate on the APD, the emission has been send directly to an APD using only a band-pass filter to spectrally filter away other emission lines in experimental Setup 2 (see Appendix A for more details). With this setup we obtain the power series shown in Fig.~\ref{fig:Sven}, which is performed on a different emitter, denoted QD2, also situated spectrally close to the M3-mode and again under pulsed excitation. The measured value of $g^{(2)}(0)$ is also shown. In Setup 2 the spectral filtering is less efficient than in Setup 1, where a grating spectrometer was used, but it does have a significantly larger throughput. As a consequence, an enhanced multi-photon contribution is observed, as is quantified in  $g^{(2)}(0)$, which increases with excitation power due to contributions from other emission lines.
Because of the weaker spectral filtering the power series on the APD does not saturate at high excitation power where background emission starts to influence the measurements. In contrast, saturation is observed when directing the emission through a spectrometer and onto a CCD, cf. Fig.~\ref{fig:Sven}. The CCD counts are scaled to the APD counts in order to make a proper comparison between the two measurements, which allows determining a count rate on the APD at saturation of $\mathrm{C^X_{sat}}=962\pm46$ kHz.
The finite value of $g^{(2)}(0)$ is attributed to other emission lines that make up $g^{(2)}(0)/2$ of the total signal~\cite{Madsen.NJP.2013} meaning that the contribution from the single QD line amounts to $722$ kHz. By measuring the transmission through every optical element we also determine the total transmission of this setup to be $\eta_\mathrm{setup}=(12.0\pm1.4)\%$. Under the idealized assumption of no blinking or polarization mixing a collection efficiency of $\eta_\mathrm{X}=(15.1\pm 2.0) \; \%$ is extracted. Interestingly this directly detected efficiency is found to be almost a factor of three times smaller than the relative efficiency discussed previously, which is considered in further detail below. The relevant quantities for both QD1 (Fig.~\ref{fig:1}) and QD2 (Fig.~\ref{fig:Sven}) are summarized in Table~\ref{table1} below together with some of the best values found in the literature.
\\
\begin{table}[h]
\begin{tabular}{|p{2.04cm}|c|c|c|c|}
\hline
                         & $\mathrm{C^X_{sat}}\; (\mathrm{kHz})$    & $g^{(2)}(0)\; (\%)$  &$\eta_\mathrm{X} \; (\%)$& $\eta_\mathrm{setup}\;(\%)$\\ \hline
  QD1 (Setup 1)   & \multicolumn{1}{|c|}{$293\pm 8.6$}   & \multicolumn{1}{|c|}{$ 4\pm5 $} & \multicolumn{1}{|c|}{$44.3\pm2.1$} & \multicolumn{1}{|c|}{$1.7\pm0.03$}\\
  QD2 (Setup 2)   & \multicolumn{1}{|c|}{$962\pm 46$ }    & \multicolumn{1}{|c|}{$50\pm1$}  & \multicolumn{1}{|c|}{$15.1\pm2.0$} & \multicolumn{1}{|c|}{$12.0\pm1.4$}\\
  Ref.~\cite{Gazzano.NatCom.2013}           & \multicolumn{1}{|c|}{$700$ }            & \multicolumn{1}{|c|}{$15$}         & \multicolumn{1}{|c|}{$79$}        & \multicolumn{1}{|c|}{$1.7$}\\
  Ref.~\cite{Claudon.NatPhot.2010}           & \multicolumn{1}{|c|}{$65$ }            & \multicolumn{1}{|c|}{$8$}         & \multicolumn{1}{|c|}{$72$}        & \multicolumn{1}{|c|}{$0.12$}\\
  Ref.~\cite{Strauf.Nat.2007}         & \multicolumn{1}{|c|}{$4000$ }            & \multicolumn{1}{|c|}{$40$}         & \multicolumn{1}{|c|}{$38$}        & \multicolumn{1}{|c|}{$13$}\\
  \hline
\end{tabular}
\caption{The figures of merit for the two investigated QDs in the two separate setups along with some of the best values reported in the literature. Note that $g^{(2)}(0)$ was measured under excitation powers of $2.1 P_\mathrm{sat}$ and $P_\mathrm{sat}$ for QD1 and QD2, respectively, while $\mathrm{C^X_{sat}}$ was obtained in the high power limit.}\label{table1}
\end{table}

Importantly, the efficiencies for QD1 and in Ref.~\cite{Claudon.NatPhot.2010} are extracted by applying different assumptions for the efficiencies than for QD2, Ref.~\cite{Gazzano.NatCom.2013}, and Ref.~\cite{Strauf.Nat.2007}. In the following we will clarify and discuss these two approximations:

$\mathbf{\boldsymbol \alpha_\mathrm{X} \boldsymbol \epsilon_\mathrm{X} / \boldsymbol \epsilon_\mathrm{bulk}=1}$: Used for QD1, where the collection efficiency is extracted by comparing the APD count rate to that of a QD in bulk GaAs. This approach assumes that the influence of blinking, charged excitons, and non-radiative decays are the same for the exciton (X) and the QD in bulk GaAs. This assumption is usually not verified directly experimentally.

$\mathbf{\boldsymbol \alpha_\mathrm{X} \boldsymbol \epsilon_\mathrm{X} =1}$: Used for QD2, where the collection efficiency is deduced directly from the counts on the APD. The collection efficiency extracted with this approach is directly relevant for applications, since it relates to the measured number of photons. From measurements we arrive at the bounds $(59\pm5)\%\leq \epsilon_\mathrm{X}\leq(72\pm6)\%$ and $\alpha_\mathrm{X}\leq1.092$, see Appendices B and C for more details. This implies that the collection efficiency is underestimated and the correct value is bound within $(19.2\pm3.0)\%\leq \eta_\mathrm{X}\leq(23.4\pm3.7)\%$, which is a conservative estimate using the upper bound on the degree of polarization mixing of $\alpha_\mathrm{X}=1.092$.

\begin{figure}[tb]
\includegraphics[width=8.3cm]{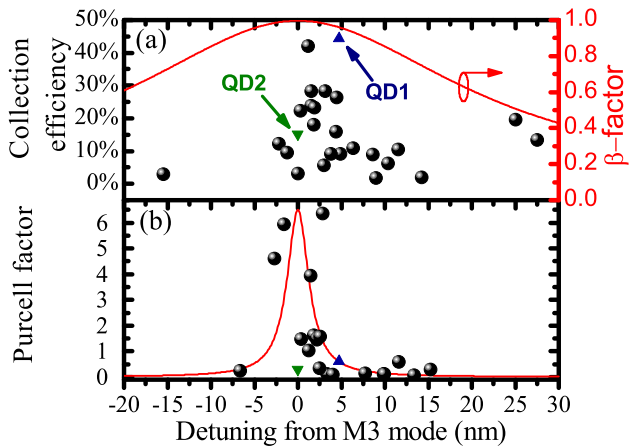}
\caption{\textbf{(a)} Recorded collection efficiency at the first lens for many QDs at different detunings (black points). The high collection efficiency is broadband and consistent with the calculated $\beta$-factor dependence (red line). \textbf{(b)} Decay rates showing Purcell enhancement when close to resonance with the cavity mode. The Purcell factors are consistent with the Lorentzian (red line) corresponding to $Q=300$, which is typical for the M3 mode.
\label{fig:det}}
\end{figure}

The collection efficiencies extracted with these two approaches differ by a factor of $2.9$. The first approach assumes $\epsilon_\mathrm{X}=\epsilon_\mathrm{bulk}$, that is the probability the QD decays by other processes than photon emission on the desired transition is identical for the two QDs. In contrast the second approach does not rely on such assumptions. We attribute the deviation between the two collection efficiencies to the spatial and spectral position of the two QDs, cf. Fig.~\ref{fig:det}(a).

A competitive benchmark of a single-photon source is whether it enables the construction of quantum gates that violate quantum locality, which puts strong bounds on the single-photon purity, indistinguishability, and overall efficiency. For instance, the construction of a nonlocal scalable two-photon "KLM-type gate"~\cite{Knill.Nature.2001} would require $g^{(2)}(0)\leq1\%$, $\eta_\mathrm{X}\geq72\%$, and $\eta_\mathrm{setup}\geq90\%$ under the assumption of unity indistinguishability and availability of photon-number resolving detectors~\cite{Jennewein}. Furthermore, for many applications not only the collection efficiency into a certain NA but rather into a single mode fiber is important. For QD2 in Setup 2 we measure that $(60\pm 5)\%$ of the emission that has been captured by the first lens is coupled into the single mode fiber. We emphasize that while these demanding requirements are not yet met, they appear to be within experimental reach with QD sources, and the system presented here is thus comparable to the best in the literature both in terms of efficiency and actual count rate on the APD, cf. Table~\ref{table1}.

We have measured the efficiency at the first lens using the assumption $\alpha_\mathrm{X}\epsilon_\mathrm{X} / \epsilon_\mathrm{bulk}=1$ for many QDs all situated spectrally around the M3 mode, cf. Fig.~\ref{fig:det}(a), and the high efficiency is found to be a broadband feature. For reference the estimated $\beta$-factor~\cite{Beta.factor} for a cavity with $Q=300$ is also plotted, and the collection efficiency qualitatively follows the $\beta$-factor defined as $\beta=\gamma_\mathrm{cav}/\gamma_\mathrm{tot}$, where $\gamma_\mathrm{cav}$ and $\gamma_\mathrm{tot}$ are the decay rates into the cavity mode and the total rate, respectively. The collection efficiency is given by $\eta_\mathrm{X}=\eta_\mathrm{cav}\beta+\eta_\mathrm{rad}(1-\beta)$, where $\eta_\mathrm{cav}$ ($\eta_\mathrm{rad}$) is the collection efficiency of the cavity mode (radiation modes), and in Fig.~\ref{fig:det}(a) we have neglected the latter and much smaller term. The Purcell factor is defined as the decay rate divided by the decay rate of a QD in bulk GaAs, i.e. $F_p=\gamma/\gamma_\mathrm{bulk}$. In Fig.~\ref{fig:det}(b) the decay rates are shown and as expected the Purcell enhancement peaks around the cavity resonance. The broadband nature of the measured collection efficiencies is in agreement with the estimated $\beta$-factor, and the Purcell factors follow the lineshape of the cavity, which confirms that the high efficiency is due to the coupling to the cavity mode.

\section{Numerical modeling}
\begin{figure*}[tb]
\includegraphics[width=13.3cm]{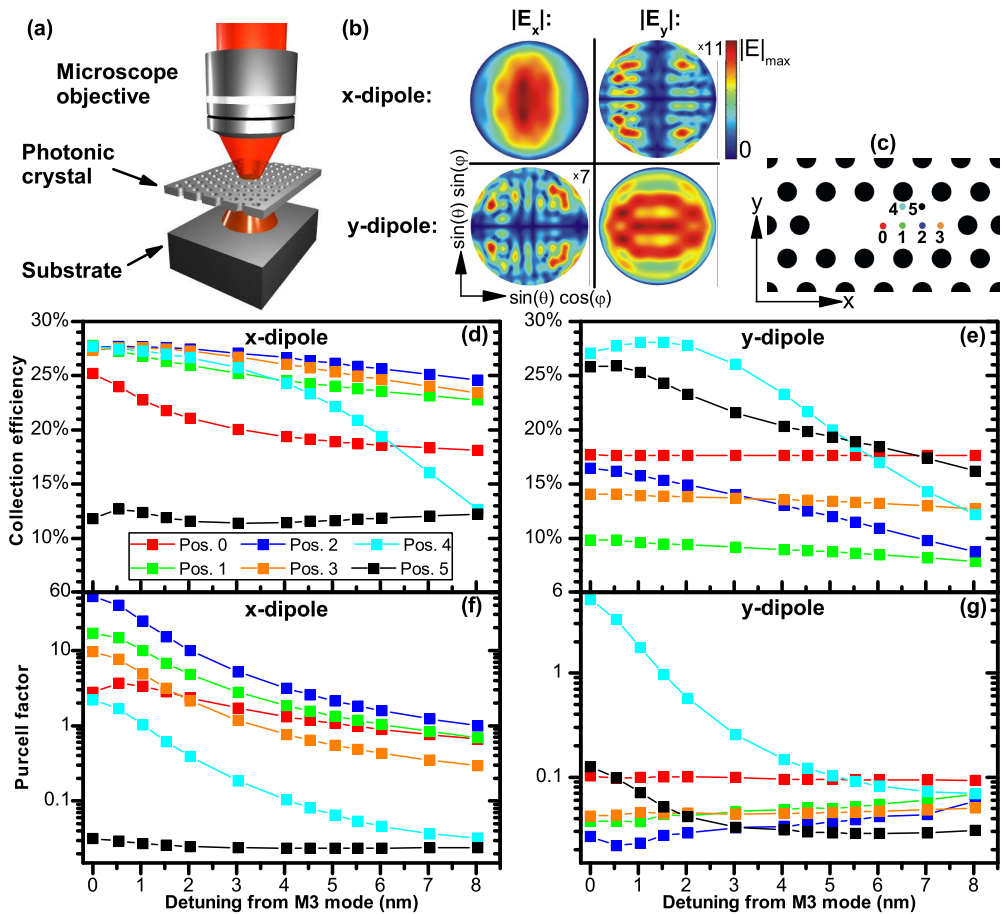}
\caption{\textbf{(a)} Sketch of the experimental situation showing the bi-directional emission from the PC membrane that is suspended over a substrate. \textbf{(b)} Calculated far-fields for the $\mathrm{x}$- and $\mathrm{y}$-dipoles in spherical coordinates with the polar ($\theta$) and azimuthal ($\varphi$) angles for position $3$ at a detuning of $5$ nm. The $\mathrm{x}$-($\mathrm{y}$-) dipole only emits into the $\mathrm{x}$-($\mathrm{y}$-) component of the far-field, which proves that there is almost no mixing of polarizations in the far field for this particular position. \textbf{(c)} Position map for the cavity. \textbf{(d)-(e)} Calculated efficiencies of $x$- and $y$-dipoles, respectively, as collected by a NA$=0.6$ objective as a function of detuning from the M3 mode for six different spatial positions in the cavity. \textbf{(f)-(g)} Calculated Purcell factors for the QD coupling to the cavity mode for many positions.
\label{fig:2}}
\end{figure*}
In order to understand the origin of the high efficiency, we have performed numerical simulations of the emission from a QD with various detunings from the M3 mode. A finite-element method is used to calculate the electric field emitted from the QD on a surface $10$ nm above the PC membrane and subsequently  perform near-field to far-field transformations over the surface~\cite{Vuckovic.IEEE.2002,Kim.PRB.2006}. In Fig.~\ref{fig:2}(a) the experimental situation is sketched, where the microscope objective collects the emission from a QD in a PC membrane suspended over the GaAs substrate. The far-field patterns of two orthogonal dipoles oriented along the $\mathrm{x}$- and $\mathrm{y}$-axis are calculated, and examples of these far-field patterns are shown in Fig.~\ref{fig:2}(b). We extract the collection efficiency of an NA$=0.6$ microscope objective corresponding to the one used experimentally, and in Fig.~\ref{fig:2}(d)-(e) plot the collection efficiency for the $\mathrm{x}$- and $\mathrm{y}$-dipoles for six different positions in the cavity as a function of detuning from the M3 cavity mode. The corresponding positions can be seen in Fig.~\ref{fig:2}(c). The $\mathrm{x}$-dipole is generally the most efficient dipole, and the efficiency remains high at large detunings for several positions. In Fig.~\ref{fig:2}(f)-(g) the calculated Purcell factor is plotted as a function of detuning for the two dipoles. This figure supports the conclusion that the high efficiency is due to coupling to the cavity mode, in good agreement with the experimental observations in Fig.~\ref{fig:det}.

As an example, position $3$ exhibits the broadband high collection efficiency and low Purcell factor that we have observed experimentally for QD1 and QD2. In Fig.~\ref{fig:2}(b), $|\mathrm{E_x}|$ and $|\mathrm{E_y}|$ are plotted in the far-field for the $\mathrm{x}$- and $\mathrm{y}$-dipoles for position $3$ at a detuning of $5$ nm. For the $\mathrm{x}$-dipole, $|\mathrm{E_y}|$ is scaled up by a factor of $11$, while $|\mathrm{E_x}|$ is scaled up by a factor of $7$ for the $\mathrm{y}$-dipole, in order to make the far fields clearly visible. We immediately observe that the $\mathrm{x}$- and $\mathrm{y}$-dipoles almost exclusively emit into the $\mathrm{x}$- and $\mathrm{y}$-polarizations in the far-field, respectively. This corresponds to $\alpha_\mathrm{X}=1$.

\begin{figure*}[t]
\includegraphics[width=13.3cm]{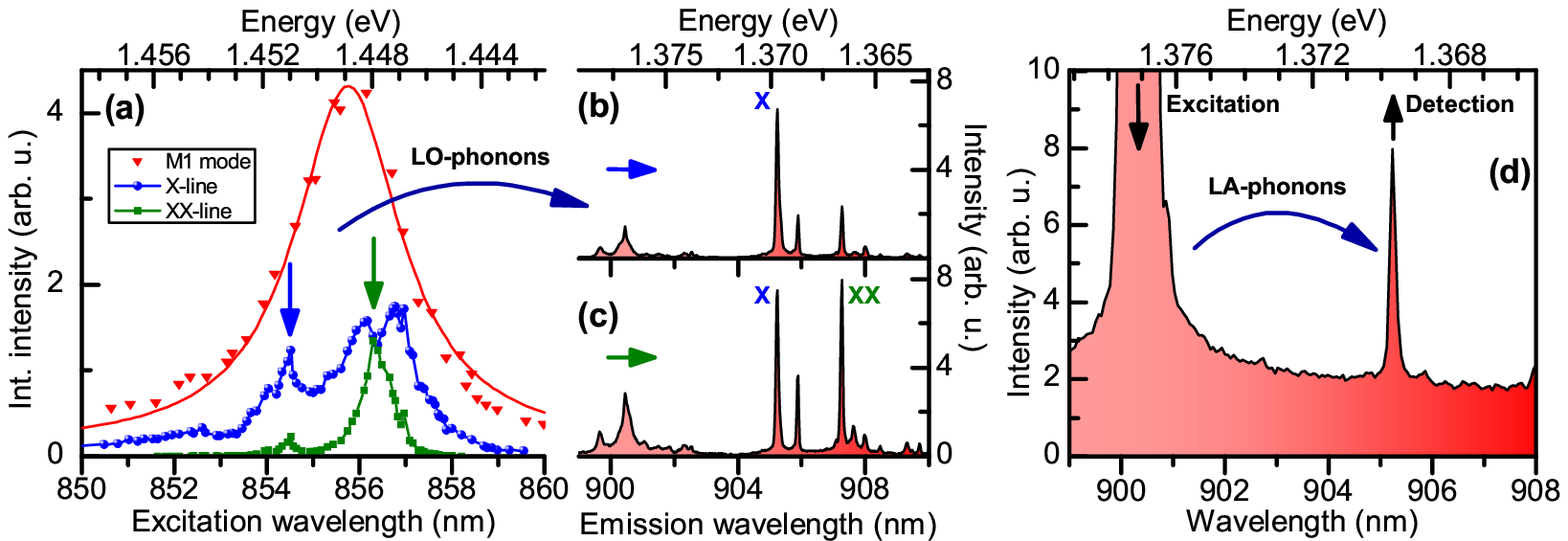}
\caption{\textbf{(a)} Integrated intensity of the M1 mode under high-power excitation and in the X- and XX-lines under low-power excitation as a function of excitation wavelength around the M6 mode. The linewidth of the M6 mode is mapped by detecting the M1 mode, while detecting the X- and XX-lines reveals sharp resonances due to the absorption of two LO-phonons. In \textbf{(b)-(c)} the resulting emission spectra when exciting at $854.56$ nm (blue arrow) and at $856.40$ nm (green arrow) are shown. \textbf{(d)} The excitation laser is tuned to the M3 mode, where LA-phonons mediate the excitation of the X-line while the X2- and XX-lines are absent.
\label{fig:phonon}}
\end{figure*}

Polarization mixing does, however, occur, e.g., for position $4$, because the $y$-dipole in this case couples to the M3 mode, which is $x$-polarized in the far-field. The strong polarization dependence of the X-line indicates that $\alpha_\mathrm{X}>1$, c.f. Appendix C, which may explain why the calculated efficiencies are systematically lower than the experimentally measured value of $\eta_\mathrm{X}=(44.3\pm2.1)\%$. Another mechanism that could increase the amount of collected light is reflections from the substrate $1530$ nm beneath the membrane that are not accounted for in the simulations.  This air-GaAs interface, cf. Fig.~\ref{fig:2}(a), will reflect $\sim55\%$ of a perpendicularly incident electric field. This reflected field can interfere constructively with the top-emitted field, and thereby increase the out-coupling efficiency significantly~\cite{Pinotsi.Thesis.2011,Kim.PRB.2006}.

\section{Phonon-mediated excitation}

In the following we address various excitation schemes of the QD leading to the generation of indistinguishable photons. For the rest of the paper only QD1 in Setup 1 is considered, and in Fig.~\ref{fig:phonon}(a) a photoluminescence excitation (PLE) measurements is shown, where the excitation laser is scanned across the M6 mode while detecting the emission intensity. First we detect and plot the total intensity in the M1 mode as a function of laser wavelength under high excitation power. In this way the equivalent of an absorption spectrum is obtained, giving a Q-factor of $306$ for the M6 mode, c.f. Fig.~\ref{fig:phonon}(a). Next the excitation laser is scanned across the M6 mode with low power, while this time recording the intensity in the X- and XX-lines. The X-line exhibits a sharp resonance at $854.56$ nm and two resonances around $856.40$ nm, while the XX-line shows a sharp resonance at $856.40$ nm. Both of these resonances lie within the linewidth of the M6 mode. The full emission spectra have also been recorded while exciting through these two resonances. Under excitation at $854.56$ nm the spectrum is very clean with only the X-line being very pronounced, cf. Fig.~\ref{fig:phonon}(b). Under excitation at $856.40$ nm the spectrum remains clean, but the XX-line has increased by a factor of $\sim3$ compared to the X-line although the excitation power is kept constant, cf. Fig.~\ref{fig:phonon}(c). This increase by a factor of $\sim3$ in the intensity of the XX-line, while excitation power and the intensity in the X-line are both constant, rules out that excitation is through a higher-order state of the QD such as a d- or f-state. The energy difference between these two absorption resonances is $3.12$ meV, which matches the difference in emission energy between the X- and XX-line. The wetting layer and LA-phonon-mediated transitions do not give rise to such sharp resonances. Absorption mediated by LO-phonons on the other hand, gives rise to sharp absorption peaks due to the discrete LO-phonon energies~\cite{Heitz.APL.1996,Steer.PRB.1996,Heitz.PRB.1997} and explain that the X-XX energy difference in absorption matches that in emission~\cite{Dufaker.PRB.2013}. Furthermore, temperature-dependent measurements (up to T=$60$ K) of the absorption spectrum show that both LO-phonon resonances shift by the same amount as the X- and XX-lines, which confirms that LO-phonons may be responsible for mediating the excitation. Although this is not an irrefutable proof, we conclude that the resonances are due to the absorption of two LO-phonons, where the shorter (longer) wavelength-resonance is LO-phonon-mediated excitation of the exciton (bi-exciton).
It should be noticed that the biexciton XX is not excited directly but rather through a two-step process, where first an exciton X is excited. This follows from the finite overlap between the X and XX lines and the broad optical pulse.
The energy difference between the excitation and emission is $81.2$ meV, corresponding to a single LO-phonon energy of $40.6$ meV. For comparison the LO-phonon energy in GaAs is calculated to $36.8$ meV, and the discrepancy between the calculated and measured value is attributed to inhomogeneity and strain within the QD. Shifts of the LO-phonon energy by a similar or larger amount have been observed experimentally~\cite{Leamitre.PRB.2001,Kowalik.APL.2007} and predicted theoretically~\cite{Vasil.PRB.2004}. Furthermore, any confinement of the LO-phonons within the QD would also shift the energy of the LO-phonon~\cite{Cardona}.
Comparing count rates under above-band and LO-phonon mediated excitation, respectively, reveals that the latter is $7$ times more efficient. The total absorption probability depends on both the probability of a QD to capture a generated electron-hole pair but also on the probability of coupling the excitation light into the cavity. The latter can be enhanced by the cavity resonances leading to the phenomenon of cavity-assisted excitation.
While the QD has absorption resonances corresponding to both 1 and 2 LO-phonon lines, the PC structure suppresses absorption at the 1 LO-phonon line because its frequency is within the band-gap, while the 2 LO-phonon line is found to match the M6 cavity mode resonance and thus strongly enhances the absorption by a factor of $\sim300$ corresponding to the Q-factor of the mode.
From the bi-exciton absorption spectrum a linewidth of $1.28$ meV of the LO phonon resonance can be extracted. For comparison the absorption linewidth is estimated to be $0.25$ meV using the theory in Ref.~\cite{Krummheuer} that includes the dispersion of the LO-phonons and the size of the exciton wavefunction, which is in reasonable agreement with the experimental value. We attribute the additional broadening observed in the experiment to be due to the short lifetime ($\sim 9$ ps~\cite{Kash.Solid.1988} corresponding to a width in energy of $0.145$ meV) of LO-phonons~\cite{Stock.PRB.2011}.

\begin{figure*}[t]
\includegraphics[width=13.3cm]{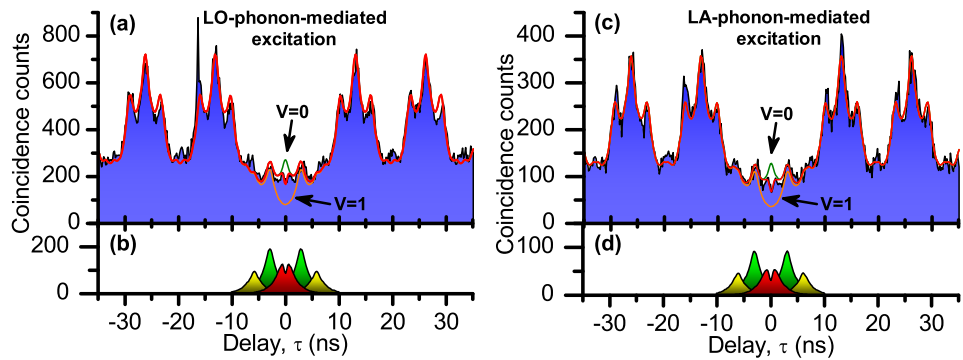}
\caption{\textbf{(a)} Indistinguishability measurements under LO-phonon-mediated excitation at $5$ K and at an excitation power of $0.88 \; \mathrm{P_{sat}}$. Modeling of the data (red curve) gives $\mathrm{V_{LO}}=(13 \pm 2)\%$ and in \textbf{(b)} the theory curve is decomposed into the five peaks, where the center peak (red area) corresponds to the desired two-photon interference. \textbf{(c)-(d)} Shows the same under LA-phonon-mediated excitation at $4.8$ K and at the excitation power $0.10 \; \mathrm{P_{sat}}$, and we extract $\mathrm{V_{LA}}=(19 \pm 4)\%$.
\label{fig:HOM}}
\end{figure*}

Figure~\ref{fig:phonon}(d) demonstrates that it is also possible to selectively excite the X-line by exciting through the nearby M3 mode. With this excitation method the X2- and XX-lines are found to be completely absent, which proves that this excitation method is quasi-resonant where the angular momentum is preserved during the phonon relaxation process. This excitation mechanism is mediated by LA-phonons that form a continuum and the residual energy of the incoming photons is emitted into the lattice as LA-phonons. Previous work has demonstrated LA-phonon emission of the residual energy in a cavity quantum electrodynamics setting for the same detunings~\cite{Madsen.arxiv.2012,Weiler.PRB.2013}. Efficient excitation through LA-phonons has also been predicted in the different regime of longer excitation pulses and shorter detunings~\cite{Glassl.PRL.2013}. In conclusion we can efficiently and selectively excite a single exciton in a QD by cavity-enhanced phonon-mediated processes, where the energy difference between excitation and emission is absorbed by either two LO-phonons or by single LA-phonons.

\section{Indistinguishability measurements}

The various phonon-mediated excitation schemes are subsequently explored as a mean to generate indistinguishable (i.e., coherent) single photons. The coherence is measured by performing two-photon interference measurements, where two consecutively emitted photons from the same QD are interfered on a beam-splitter in a Hong-Ou-Mandel (HOM) interferometer, c.f. Appendix D for more details. The outcome of such a HOM measurement, cf. Fig.~\ref{fig:HOM}(a)-(c), gives a cluster of $5$ peaks for every $13$ ns, where only the center peak corresponds to the desired two-photon interference. The coincidence counts in the center peak are given by~\cite{Kiraz.PRA.2004}
\begin{equation}\label{HOMeq}
G^{(2)}_\mathrm{HOM}(\tau)=\frac{1}{4\gamma} e^{-\gamma |\tau|}(1-e^{-2\gamma_\mathrm{dp}|\tau|})\;\;,
\end{equation}
where $\tau$ is the time delay, $\gamma$ the decay rate, and $\gamma_{\mathrm{dp}}$ is the pure-dephasing rate included to model the effect of decoherence caused by the solid-state environment. This expression holds for the ideal case of a pure single-photon source ($g^{(2)}(0) \sim 0$) as verified in Fig.~\ref{fig:1}(e), cf. discussion in Appendix D. Equation~(\ref{HOMeq}) shows that in the absence of dephasing ($\gamma_{\mathrm{dp}}=0$) the photons are completely indistinguishable and the center peak vanishes. Comparing the area in the center peak, S$_0$, where two photons interfere, with the area in the neighboring peak, S$_1$, where the two photons are separated by $3.04$ ns and consequently do not interfere, allows us to express the degree of indistinguishability as $\mathrm{V}=1-S_0/S_1=\gamma/ (\gamma+2\gamma_{\mathrm{dp}})$.

In Fig.~\ref{fig:HOM}(a) the outcome of a HOM measurement on the X-line under LO-phonon-mediated excitation is shown. The theory curve shown in Fig.~\ref{fig:HOM}(a) is obtained using the procedure described in Appendix D, and we extract an indistinguishability of $\mathrm{V_{LO}}=(13 \pm 2)\%$ under LO-phonon-mediated excitation. For reference, the curves with $\mathrm{V}=0$ and $\mathrm{V}=1$ are also shown, and in Fig.~\ref{fig:HOM}(b) the model is decomposed into the five peaks for clarity. It is observed that the central peak has the structure of Eq.~(\ref{HOMeq}), where the dephasing rate gives the width of the dip. We note that even for a degree of indistinguishability of unity $(V=1)$ the central peak would not vanish fully, which is a consequence of the slow decay time of the investigated QD giving rise to overlapping peaks. From the indistinguishability we extract the decoherence rate $\hbar(\gamma/2+\gamma^{\mathrm{\;LO}}_{\mathrm{dp}})=1.53\pm 0.25\;\micro$eV, which gives a pure dephasing rate of $\hbar\gamma^{\mathrm{\;LO}}_{\mathrm{dp}}=1.33 \pm 0.25\;\micro$eV at a temperature of $5$ K and at an excitation power of $0.88\;\mathrm{P_{sat}}$.

Similar measurements and data analysis are subsequently performed on the X-line under LA-phonon-mediated excitation, and the results are shown in Fig.~\ref{fig:HOM}(c)-(d). We obtain an indistinguishability degree of $\mathrm{V_{LA}}=(19 \pm 4)\%$, which corresponds to a decoherence rate of $\hbar(\gamma/2+\gamma^{\mathrm{\;LA}}_{\mathrm{dp}})=1.05 \pm 0.21\;\micro$eV and a pure-dephasing rate of $\hbar\gamma^{\mathrm{\;LA}}_{\mathrm{dp}}=0.85 \pm 0.21\;\micro$eV at a temperature of $4.8$ K and at an excitation power of $0.10 \; \mathrm{P_{sat}}$. The results are summarized in Table~\ref{table2}, where $T_1=1/\gamma$ is the lifetime, $T_2^*=1/\gamma_\mathrm{dp}$ is the pure-dephasing time, and $T_2$ is the coherence time defined as $\frac{1}{T_2}=\frac{1}{2T_1}+\frac{1}{T_2^*}$:
\\
\begin{table}[h]
\begin{tabular}{|p{3.1cm}|p{2.5cm}|p{2.5cm}|}
  \hline
   & LO-phonon-mediated exc. & LA-phonon-mediated exc. \\
  \hline
  $\mathrm{T}_1=1/\gamma$ & \multicolumn{1}{|c|}{1.61 ns}  & \multicolumn{1}{|c|}{1.61 ns}   \\
  $\mathrm{T}_2^*=1/\gamma_\mathrm{dp}$  & \multicolumn{1}{|c|}{$0.49\pm0.09$ ns} & \multicolumn{1}{|c|}{$0.77\pm 0.19$ ns}                   \\
  $\mathrm{T}_2=\left( \frac{1}{2T_1}+\frac{1}{T_2^*}\right)^{-1}$    & \multicolumn{1}{|c|}{$0.43\pm0.07$ ns} & \multicolumn{1}{|c|}{$0.63\pm0.13$ ns}                   \\
  \hline
  \end{tabular}
  \caption{The lifetimes, pure-dephasing times, and coherence times under LO- and LA-phonon mediated excitation.}\label{table2}
  \end{table}

Changing the excitation from LO- to LA-phonon-mediated excitation improves the indistinguishability, which we partly attribute to the significantly lower excitation power and partly to the fact that the residual energy that is emitted as LO- and LA-phonons, respectively, is much smaller when exciting through LA-phonons. For the LO-phonon (LA-phonon) excitation the relevant phonon emission energy is $81.2$ meV ($7.2$ meV), cf. Fig.~\ref{fig:phonon}.
The corresponding thermal occupation of LO- and LA-phonons at these energies is effectively negligible at 5 K, and therefore the spontaneously emitted phonon in both cases would be expected to constitute a significant perturbation of the system. Furthermore, if both dipoles contribute to the signal, i.e., $\alpha_\mathrm{X}>1$, the frequency difference between the two dipoles will give rise to a beating. In the measurement this will appear as a degradation of the visibility and thus contribute to the extracted dephasing rate. Although the measured degrees of indistinguishability are not very high, the pure dephasing times are rather long when comparing to the, to our knowledge, only previous measurement on a QD in a PC cavity, where a value of $T^*_2=0.281$ ns was measured~\cite{Laurent.APL.2005}.
Longer pure-dephasing times of up to $T^*_2=5.7$ ns have been measured on a QD in bulk GaAs under resonant excitation~\cite{He.NatNan.2013}, and on Purcell enhanced QDs in micropillar cavities~\cite{Gazzano.NatCom.2013,Santori.Nature.2002}. However, we stress the difference that our measurements are done on a QD embedded in a nanostructure, thus experiencing a highly inhomogeneous solid-state environment. For the purpose of generating indistinguishable photons from QDs in PCs we calculate that an indistinguishability of $V=70\%$ can be achieved from a QD with a moderate Purcell factor of $6$, which is the level we have measured experimentally in Fig.~\ref{fig:det}(b). Furthermore, indistinguishabilities of $V=78\%$ and $V=85 \%$ can be achieved using the highest measured Purcell factors of 9 and 15 for QDs in PC waveguides and PC cavities, respectively~\cite{Lodahl.arXiv.2013}. These numbers highlight the prospect of generating highly indistinguishable photons from QD in PC nanostructures. We emphasize that these numbers are expected to improve even further when implementing strict resonant excitation.
Furthermore we stress, that previous experiments ~\cite{Santori.Nature.2002,Gazzano.NatCom.2013} have focussed on Purcell-enhanced QDs allowing them only to investigate the integrated area of the center peak, whereas here the dip within the central peak is resolved due to the inhibition of the decay rate caused by our PC structure. This dip occurs because the time intervals between photodetections are shorter than the mutual coherence time of the two photons, thus making them indistinguishable, and this effect has been observed previously in atomic optics~\cite{Legero.PRL.2004}, but never before for QDs.  In conclusion this proves that in addition to being a very well-suited platform for planar integration, PCs can also be used as a platform for generating coherent single photons, despite the strong structural inhomogeneity.

\section{Concluding remarks}
We have presented measurements on a QD detuned from a low-Q mode of a PC cavity with a collection efficiency at the first lens of $\eta_\mathrm{X}=(44.3\pm2.1)$\%, while maintaining single-photon emission. Experimentally and numerically the high collection efficiency is found to be a broadband feature, originating from the high and broadband $\beta$-factor, which is a consequence of the photonic bandgap. Count rates as high as $\mathrm{C^X_{sat}}=962\pm46$ kHz have been achieved, which makes this system a promising candidate for the application in quantum computing. We have demonstrated LO-phonon-mediated excitation of both the exciton and bi-exciton by the absorption of two LO-phonons as well as LA-phonon-mediated excitation. HOM measurements under LO- and LA-phonon-mediated excitation showed very low dephasing rates. The recorded low dephasing rates in a PC structure are very important for the prospects of using PCs as a platform for quantum-information processing. Furthermore, the inhibition of the decay rate has allowed us to resolve the central dip in the HOM measurements, which was first observed for photons emitted from atoms.

\subsection{Acknowledgments}
We thank Niels Gregersen for simulations of the QD in bulk GaAs, Yuntian Chen for fruitful discussions on the finite-element simulations, and we gratefully acknowledge financial support from the Villum Foundation, the Carlsberg Foundation, the Danish Council for Independent Research (Natural Sciences and Technology and Production Sciences) and the European Research Council (ERC consolidator grant 'ALLQUANTUM').

\appendix

\section{Appendix A: Experimental details}
\textbf{Sample:}
The sample is a GaAs PC membrane with lattice constant $a=240$ nm, hole radius $r=66$ nm, and thickness $154$ nm. The sample contains self-assembled InGaAs QDs embedded in the center of the membrane with a density of $80\;\micro$m$^{-2}$. An L3 cavity is introduced by leaving out three holes on a row and the Q-factor of the first-order mode is optimized by shifting the three holes at the ends of the cavity by $0.175a$, $0.025a$, and $0.175a$, respectively~\cite{Akahane.OE.2005}.
\\

\textbf{Setup 1:}
The sample is placed in a He-flow cryostat and probed optically by confocal microscopy using a microscope objective with numerical aperture NA=$0.6$. A dichroic mirror (cut-off at $900$ nm) is used to separate the emission from the excitation laser. A tunable Ti:Sapph laser, that can operate both in CW and pulsed ($3$ ps pulse duration and $76$ MHz repetition rate), is used for excitation of the sample. The emission passes through a half-wave plate and a polarizing beam-splitter for polarization filtering, and afterwards it is sent to a spectrometer through either a free-space path (for measurements of the efficiency) or through a single-mode polarization-maintaining fiber (for autocorrelation and indistinguishability measurements). After the spectrometer, the emission is directed onto a CCD-camera or an avalanche photodiode (APD).
\\

\textbf{Setup 2:}
In order to achieve the highest possible count rate on the APD we also use a setup where we excite the sample with another tunable Ti:Sapph laser (3 ps pulse duration and $80$ MHz repetition rate), collect the emission with a NA$=0.85$ microscope objective, and couple the emission into a polarization-maintaining fiber. We only use a band-pass filter to spectrally filter away other emission lines before directing the emission onto an APD.

\section{Appendix B: Preparation efficiency}

The preparation efficiency $\epsilon$ specifies the number of photons emitted from the neutral excitons ($x$- and $y$-dipoles) per excitation pulse. $\epsilon$ can be reduced below unity for several reasons: the QD can go to a dark state known as blinking, non-radiative decay processes give rise to a quantum efficiency below unity, and finally not only neutral excitons but also charged excitons can be prepared in the QD. In the following we address these three issues.

Firstly, we have investigated autocorrelation measurements on a timescale of up to 10 milliseconds that show no sign of blinking. For excitation powers below and far above saturation the standard deviation of the peak heights are $\pm2.5\%$ and $\pm 1.6\%$, respectively.

Secondly, due to the presence of a dark state the decay of the neutral exciton is bi-exponential and the internal quantum efficiency ($\eta_\mathrm{QE}$) is defined as $\eta_\mathrm{QE}=(\gamma_\mathrm{fast}-\gamma_\mathrm{nrad})/\gamma_\mathrm{fast}$, where $\gamma_\mathrm{fast}$ is the fast decay rate of the bi-exponential decay and $\gamma_\mathrm{nrad}$ is the non-radiative decay rate. Here we have neglected the spin-flip rate because it is typically an order of magnitude smaller than $\gamma_\mathrm{nrad}$. For a QD in bulk GaAs, the non-radiative rate is given by the slow decay rate of the bi-exponential decay. However, we observe that the slow decay rate is much faster than expected from measurements on QDs in bulk GaAs or in a PC without an L3 cavity~\cite{Wang.PRL.2011}. This indicates that polarization mixing is present, which gives rise to a contribution to the slow decay rate originating from the radiative decay of the orthogonal $y$-dipole. Therefore, in order to estimate the quantum efficiency we use the average non-radiative decay rate $\gamma_\mathrm{nrad}=0.06\pm0.05\;\mathrm{ns^{-1}}$ measured in Ref.~\cite{Wang.PRL.2011} on a QD in bulk GaAs, which gives $\eta_\mathrm{QE}=(90\pm8)\%$.

Finally, sometimes a charged exciton is formed, and from Fig 1(b) in the main text we extract the ratio $\mathrm{I_{X2}}/\mathrm{I_{X}}=0.52$, where $\mathrm{I_{X2}}$ and $\mathrm{I_{X}}$ are the intensities in the X2- and X-lines respectively. While the X-line contains two orthogonal linear dipoles split in energy by the fine-structure splitting, the X2-line contains two orthogonal circular dipoles that are degenerate in energy~\cite{Bayer.PRB}. The measured intensity ratio is related to the ratio of initial population of the X2 and X-state by the following equation:
\begin{align}
\frac{\mathrm{I_{X2}}}{\mathrm{I_{X}}} =\frac{\xi_\mathrm{X2}}{\xi_\mathrm{X}} \left( \frac{\beta^\mathrm{R}+\beta^\mathrm{L}}{\beta^\mathrm{x}+\beta^\mathrm{y}}\right)\;\; , \;\;\; 1\leq \frac{\beta^\mathrm{R}+\beta^\mathrm{L}}{\beta^\mathrm{x}+\beta^\mathrm{y}}\leq2 \;\;,
\end{align}
where $\xi_\mathrm{X}$ ($\xi_\mathrm{X2}$) is the average initial population of the X (X2) states. $\beta^\mathrm{i}$ is the emission into the cavity mode divided by the total emission from an $i$-dipole, where R and L (x and y) denote the two orthogonal circularly (linearly) polarized dipoles. We note that the cavity field is found always to be linearly polarized. The two extrema correspond to the field, at the QD position, being aligned along the x-axis or $45$ degrees with respect to it, which gives the upper bound of 2 and lower bound of 1. Formally these bounds only hold under the assumption that the decay rate into leaky modes is the same for the $x$- and $y$-dipoles, but they do also hold when these rates are much smaller than the decay rate into the cavity mode, which numerical simulations confirm is often the case. This allows us to obtain the bound $0.26\leq\xi_\mathrm{X2}/\xi_\mathrm{X}\leq0.52$ on the initial populations. Note that we have assumed the same $\eta_\mathrm{QE}$ for the X- and X2-line.

By taking the ratio $\xi_\mathrm{X} /(\xi_\mathrm{X}+\xi_\mathrm{X2})$ we obtain that the neutral exciton is initially populated between 0.66 and 0.79 times per excitation pulse. By multiplying with the $\eta_\mathrm{QE}$ extracted above, we obtain the following bounds on the preparation efficiency
\begin{align}
(59\pm5)\%\leq \epsilon_\mathrm{X}\leq(72\pm6)\%\;.
\end{align}

Unfortunately it was not possible to perform the same analysis on the QD in bulk GaAs, because we were not able to identify the charged exciton. However, from the biexponential decay it is still possible to extract the quantum efficiency, which gives the following upper bound for the QD in bulk GaAs

\begin{align}
\epsilon_\mathrm{bulk} \leq (94 \pm 4)\%\;.
\end{align}

\section{Appendix C: Far-field polarization mixing}

\begin{figure}[tb]
\includegraphics[width=8.3cm]{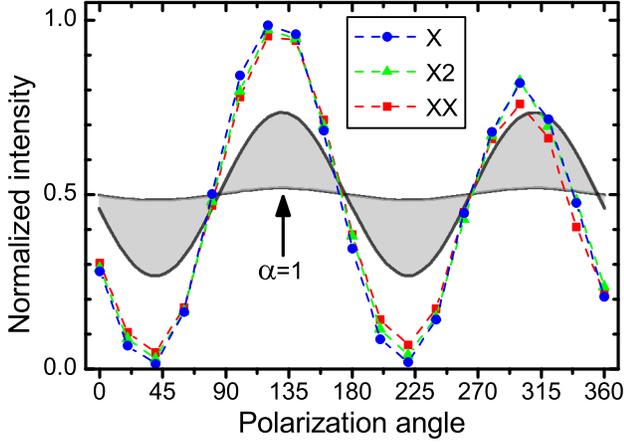}
\caption{Measured intensity, $I_x/(I_x+I_y)$, in the neutral exciton (X), charged exciton (X2), and bi-exciton (XX) lines as a function of polarization angle. All three lines are strongly polarized yielding $\rho=0.962$, 0.925, and 0.870, respectively. Shaded area contains the polarization dependencies obtained from the simulations for the positions 0-3 showing no polarization mixing, i.e. $\alpha=1$. This clearly shows that a polarization dependence is expected even when $\alpha=1$.\label{fig:S1}}
\end{figure}

Far-field polarization mixing occurs when the emission from both the $x$- and $y$-dipoles is $x$-polarized in the far-field. The M3 mode has an $x$-polarized far-field, and a strongly detuning-dependent Purcell factor is the sign of coupling to the cavity. From Fig. 4(f) in the main text, it is clear that the $x$-dipoles couple to the cavity mode for all position except 4, while only positions 4 and 5 show $y$-dipoles that couple to the cavity mode. We thus conclude that only position 4 shows polarization mixing. In order to compare the simulations to experiments we first define the fraction, $\rho$, of $x$-polarized intensity to the total intensity
\begin{align}\nonumber
\rho&=\frac{I_x}{I_x+I_y}=\frac{(\beta_{yx}+\beta_{xx})\eta_x}{(\beta_{yx}+\beta_{xx})\eta_x+(\beta_{yy}+\beta_{xy})\eta_y} \\ &=\left(1+\frac{\eta_y}{\eta_x}\left( \frac{2-\alpha}{\alpha}\right)\right) ^{-1}\; , \label{EQ1} \\
\alpha&=1+\beta_{xx}-\beta_{yy}\; ,
\end{align}
where $\beta_\mathrm{ij}$ denotes the coupling of the i-dipole into the j-polarized far-field. $\mathrm{I_x}$ and $\mathrm{I_y}$ denote the intensities in the $x$- and $y$-polarized far-fields that are detected with the efficiencies $\eta_\mathrm{x}$ and $\eta_\mathrm{y}$, respectively. $\alpha$ denotes the polarization mixing, where $\alpha=1$ denotes no mixing and $\alpha=0$ ($\alpha=2$) denotes that both dipoles emit exclusively into the $y$($x$)-polarized far-field. For a QD in bulk GaAs we observe $\rho \simeq0.5$, which is in good agreement with the fact that $\alpha=1$ for a QD in bulk GaAs. However, Eq.~(\ref{EQ1}) shows that $\alpha=1$ does not imply $\rho=0.5$ since in general in a nanostructure $\eta_\mathrm{x}\neq\eta_\mathrm{y}$. As an example, the simulation results for position 1 show $\alpha=1$ while $\rho=0.733$. In Fig.~\ref{fig:S1} we have plotted the normalized intensity as a function of polarization angle, and the shaded area contains the positions 0-3 that all show $\alpha=1$, because all these positions are on a symmetry line where $\mathrm{E_y}=0$.

In Fig.~\ref{fig:S1} we also show the measured intensity of the X-, X2-, and XX-lines as a function of polarization angle. All three lines completely follow the polarization of the M3 mode, and we extract $\rho$ to be 0.962, 0.925, and 0.870 for the X-, X2-, and XX-lines respectively. The X2-line is strongly polarized although the contributing dipoles are circularly polarized, which highlights that the far-field polarization does not directly correspond to the near-field polarization. The large value of $\rho$ for all three lines indicate that $\alpha_\mathrm{X}>1$, which is further supported by the fact that all three lines are much stronger polarized than expected from simulations. However, we do note that the reflection from the substrate can greatly enhance, e.g., $\eta_\mathrm{X}$ and this can give rise to a high value of $\rho$, while maintaining $\alpha_\mathrm{X}=1$.

Finally, the portion of the emission from the $y$-dipole, which is $x$-polarized in the far-field, will exhibit a slower decay rate than the emission from the $x-$dipole. The recorded decay curve is bi-exponential, and assuming the slow decay to originate solely from the radiative decay of the $y$-dipole, we can express the time-dependent intensity as $I(t)=A_\mathrm{fast}e^{-\gamma_\mathrm{fast}t}+A_\mathrm{slow}e^{-\gamma_\mathrm{slow}t}$. The integrated intensity is thus given by $I=I_\mathrm{fast}+I_\mathrm{slow}$, where $I_\mathrm{slow}= \alpha_\mathrm{X}-1$ under the assumption $\alpha_\mathrm{X}\geq1$. From the recorded decay curve we extract $I_\mathrm{slow}/I_\mathrm{fast}=0.092$, which results in the following upper bound, $\alpha_\mathrm{X}\leq1.092$.

In conclusion we find strong indications that $\alpha_\mathrm{X}>1$, but we also deduce the upper bound $\alpha_\mathrm{X}\leq1.092$.

\section{Appendix D: Modeling of indistinguishability measurements}

\begin{figure*}[htb]
\includegraphics[width=13.3cm]{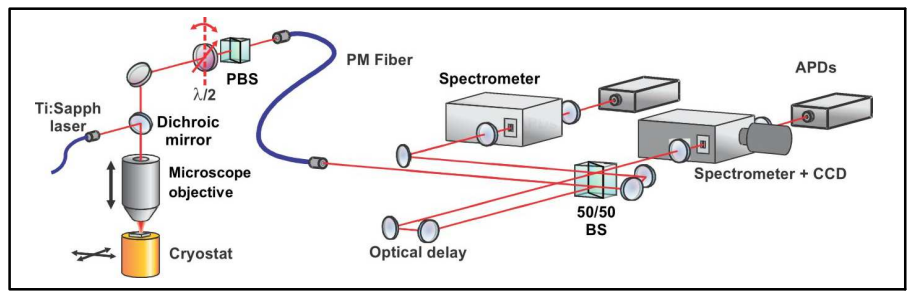}
\caption{ Experimental setup for the HOM-measurements.
\label{fig:S2}}
\end{figure*}

We measure the indistinguishability of consecutively emitted photons by interfering them in a Hong-Ou-Mandel (HOM) interferometer. For these measurements, pulsed excitation is used, where every $13$ ns two pulses separated by $3.04$ ns excite the sample. Ideally, each excitation event will give rise to the emission of a single photon, and as shown in Fig.~\ref{fig:S2} these are incident on a $50$:$50$ beam-splitter, where one arm has a delay corresponding to $3.04$ ns compared to the other arm. Both arms reflect the beam back to the same beam-splitter for interference and the coincidence counts are detected on APDs.

From the data in Fig. 6 in the main text we immediately notice that the five peaks within one cluster all overlap and the coincidence counts never reach zero. The reason is that each peak decays exponentially with the exciton decay rate and as described earlier this rate is inhibited giving rise to the large peak overlaps. The coincidence counts in the central cluster are given by the expression
\begin{align} \nonumber
G^{(2)}_\mathrm{HOM}(\tau)= &A e^{-\gamma |\tau|}(1-e^{-2\gamma_\mathrm{dp} |\tau|})\\ \nonumber
+&A( e^{-\gamma |\tau-\delta|}+e^{-\gamma |\tau+\delta|})\\
+&A/2( e^{-\gamma |\tau-2\delta|}+e^{-\gamma |\tau+2\delta|}) \;, \label{HOMeq_full}
\end{align}
where $\gamma_{\mathrm{dp}}$ is the dephasing rate, $\delta=3.04$ ns is the delay time in the interferometer, and $A$ the amplitude corresponding to the coincidence count rate on the APD. From Eq.~(\ref{HOMeq_full}) it is thus clear that for the central cluster of peaks, the peak ratios are given by $1:2:2:2:1$. However, for all other clusters of peaks, the peak ratio is expected to be $1:4:6:4:1$, because it contains correlations with both the previous and following pulses. Thus Eq.~(\ref{HOMeq_full}) can still be used but with the amplitudes corrected, by taking the limit $\gamma_\mathrm{dp}\rightarrow \infty$, and by shifting the time axis by $13$ ns corresponding to the repetition time of the laser.
In modeling the experimental data, we initially leave out the data from the central cluster of peaks. All the neighboring clusters are thus modeled with the expression in Eq.~(\ref{HOMeq_full}) modified as just described.
Secondly, we include the central cluster of peaks, but leave out the area around the central peak, and in this way we obtain the amplitude $A$. We have independently measured the time delay $\delta$ between the peaks through time-resolved measurements.
Finally, we model the central cluster of peaks with Eq.~(\ref{HOMeq_full}), where $\gamma_{\mathrm{dp}}$ is the only free parameters. From the dephasing and decay rates ($\gamma_{\mathrm{dp}}$ and $\gamma$) the visibility is easily extracted as $V=\gamma/ (\gamma+2\gamma_{\mathrm{dp}})$. Before modeling, all functions are convoluted with the instrument response function (IRF) of the setup, which is measured by sending a laser pulse through the detection setup with the delay arm blocked.


\begin{thebibliography}{16}
\bibitem{Beta.factor} The $\beta$-factor is estimated by modeling the detuning dependence of the Purcell factors in Fig.~\ref{fig:2}(f) with the expression $\mathrm{F_p=F_p^{rad}+F_p^{cav}(\Delta)}$, where $F_p^{cav}(\Delta)$ is the Purcell factor of coupling into the cavity mode that has a lorentzian lineshape. $F_p^{rad}$ is the Purcell factor for coupling to radiation modes that is assumed to be detuning independent. From the measurement at large detuning we obtain  $F_p^{rad} \sim 0.1$. The $\beta$-factor is obtained as $\beta=F_p^{cav}(\Delta)/F_p$.
\end{thebibliography}
\end{document}